\documentclass[aps,preprint,showpacs,superscriptaddress]{revtex4}
\usepackage{amssymb}
\usepackage{bm}
\usepackage{graphicx}
\def \b{\begin{equation}}
\def \e{\end{equation}}
\def \ba{\begin{array}}
\def \ea{\end{array}}
\def \be{\begin{eqnarray}}
\def \ee{\end{eqnarray}}

\def \l{\left}
\def \r{\right}
\begin{document}
\title
{Baryon kinetic energy loss in the color flux tube model.}
\author{K.A. Lyakhov}
\affiliation{Frankfurt International Graduate School for Science, J.W. Goethe Universit\"{a}t,\\
Max-von-Laue Str. 1, D--60438 Frankfurt am Main, Germany}

\begin{abstract}
This article generalizes Schwinger's mechanism for particles production in
the arbitrary finite field volume. McLerran-Venugopolan(MV) model and iterative solution of DGLAP equation in the double leading log approximation for small x gluon distribution function were used to derive the new formula for initial chromofield energy density. This initial chromofield energy is distributed among color neutral clusters or strings of different length. This strings are stretched by receding nucleus. From the proposed mechanism of string fragmentation or color field decay based on exact solution of Dirac equation in the different finite volume, the new formulae for esimated baryon kinetic energy loss and rapidity spectrum of produced partons were derived.

\end{abstract}

\pacs{12.38.Mh, 25.75.Nq, 25.75.-q} \maketitle

\section{Introduction}
                                                                                                                                                                                                                                                                 
The process of partons production can be considered as the tunnelling across the energetic gap of width $2m_{\perp}$ between the virtual energetic states inside Dirac sea to the real states--produced pairs with nonzero momentum. Semiclassical or WKB consideration of pair production was performed in \cite{popov}, \cite{mat} and in many others. This article generalizes the Schwinger's mechanism for particles production in infinite field volume \cite{sch} for the case when caps of color flux tube recede from each other quasistatically. We have used low $x$ behavior of parton distribution function to calculate dispersion of color charge per unit area and hence initial chromofield energy density. Moreover it will be shown how the total probability of string decay is related to the rapidity spectrum of produced partons. 

The current work uses the same methods as in \cite{wang}, where finite field volume effects were considered first, and improves their results. Finite size effects on pair production in transverse direction were taken into account by applying MIT boundary conditions \cite{wilets}. The other possible but less general way to introduce influence of finite size effects on particles production is based on further development of Green functions method. For instance in \cite{mart} finite size effects were incorporated by expansion of Green functions on inverse volume occupied by field. 

At RHIC energies ($\gamma\sim 100$) nucleus can be represented as two massive sheets leaving the strong gluon field in their wake. It is instructive to split evolution of the system produced in heavy ion collisions on three characteristic stages. On the first stage immediately after collision the nuclei because of multiple soft gluons exchange acquire stochastic color charge. 
This color charges produce multitude of color flux tubes occupying space between receding streaks. Shortly afterwards flux tubes decay on prompt partons. In the second stage they lose part of their energy due to gluon radiation in cascade. And finally in the third stage secondary rescatterings drive the system to local thermal equilibrium. 

The subject of interest of this article will be the first stage of reaction where the screening of the color field (back reaction of produced plasma) may be disregarded. Parton distribution function calculated on this very initial stage can serve as initial condition for parton cascade model (PCM) \cite{geiger}, where particles are still energetic enough to apply pQCD for calculation of collision integrals and system can be treated as classical.    

In this paper probability for each vertex is calculated by solving wave equations where the volume occupied by field is restricted in transverse direction by the MIT boundary conditions and in the longitudinal direction by the distance $L$ between colliding nuclei or capacitor plates \cite{wilets}.

\section{The model }

\begin{figure}
\includegraphics[width=12.cm,height=6.cm]{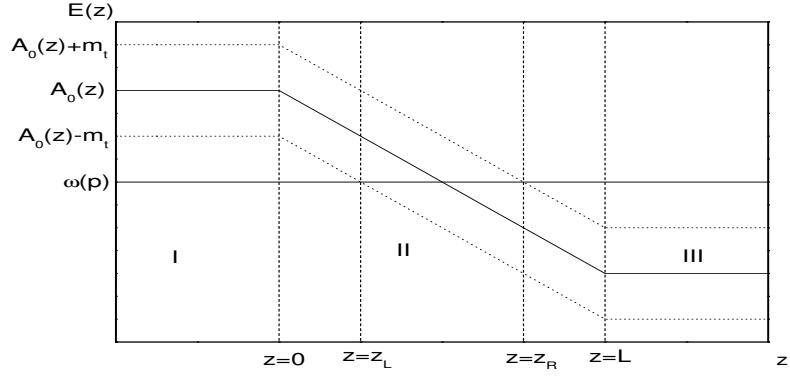}
\caption{Energetic gap between positive and negative continuum in the presence of external field (linear string potential) as a function of coordinate.} \label{fig1}
\end{figure}
QCD Lagrangian and corresponding field equations read: \b
{\cal L}_{QCD}(x)=\bar\Psi(x)[i\gamma^{\mu}{\cal D}_{\mu}-m]\Psi(x)-\frac1{4}F_{\mu\nu}^aF^{\mu\nu,a}\quad{\mbox{(sum over repeated indices) $a=1\ldots 8$}} 
\e \b \label{1} [i\gamma^{\mu}{\cal D}_{\mu}-m]\Psi(x)=0; \e 
\b \label{2} {\cal D}_{\mu}F_{\mu\nu}^a=\bar{\Psi}(x)\gamma_{\nu}\frac{\lambda^a}{2}\Psi; \e
\b{\cal D}_{\mu}=\partial_{\mu}-ig_s\frac{\lambda^a}{2}A_{\mu}^a. \e 
Gluon field is assumed gradually converting to plasma. This external field corresponds to the string potential $g_sA_{\mu}=(A_0(z),0)$. The energy stored in the color flux tube or gluon bag is regulated by its length $L$ and its radius $r_0$, see below. Thus linear vector potential is given by  
\b
A_0=\l\{\ba[c]{l}0\qquad{\mbox{for $z\le0$ (region I)}}\\
-\sigma z\qquad{\mbox{for $0\le z\le L$ (region II)}} \\
\mbox{$-\sigma L\quad$for$\quad z\ge
L\quad$(region III)}\ea \r.
\e 
which is shown on Fig.\ref{fig1}.
   
In MIT bag model the energy stored in color flux tube is constrained by its volume ${\cal A}L$, where ${\cal A}=\pi r_0^2$ is the string cross section. Therefore energy per unit length or string tension $\sigma$ can be expressed though the profile of the initial chromofield energy density $\epsilon_f(b,s)$ which will be estimated below: 
\b\sigma=(\epsilon_f(b,s)+B){\cal A},\e where $B=\Lambda_{QCD}^4$ is the bag constant, and $\Lambda_{QCD}=200 MeV$. As it is shown on Fig.\ref{fig2} the larger color flux tube radius the larger density of probability of pair production. This is easily explained by increasing string tension and therefore density of probability of string decay. 

Let us calculate initial chromofield energy density by following the ideas formulated in Refs. \cite{McL,Kov} and widely used now as the Color Glass Condensate initial state. Within this picture, random color charges are generated on the nuclear sheets as a result of soft gluon exchange at the interpenetration stage. In a single event these charges fluctuate from point to point in the transverse plane. The charges also fluctuate from event to event, so that in average the areal charge is zero. It is convinient to introduce the color charge density, $\rho(x^-,{\bf s})$, as a function of coordinate in transverse plane and light cone variable $x^-=(t-z)/\sqrt(2)$. Following Refs.\cite{McL,Kov} we assume it as a stochastic variable distributed with Gaussian weight:
\b
P[\rho]\sim\exp\l[-\int d\eta d^2{\bf s}\frac1{2\mu_a^2(\eta,{\bf s})}Tr\rho^2(\eta,{\bf s})\r],
\label{MV}
\e  
where we have applied transformation of color charge density from light cone coordinate to pseudorapidity $\eta=\ln(1/x^-)$. In contrast to CGC model dependence on $x^-$ is not important in MV model, and $x^-$ dependence is reasonable to integrate out. Therefore areal charge density 
\b
\rho({\bf s})=\int d\eta\rho(\eta,{\bf s})
\e
is random variable with gausssian distribution
\b
P[\rho]\sim\exp\l(-\frac1{2\mu_a^2}\int d^2{\bf s}\r),
\e   
Variance of color charge fluctuations in the nucleus $a$ is denoted by $\mu_a^2$. These fluctuations are characterized by a certain scale in the transverse plane, which is related to the saturation scale $a\approx\frac1{Q_s}$ introduced in high-density QCD \cite{Grib}. For central Au+Au collisions at RHIC energies $Q_s^2\approx 1.44 GeV^2$ \cite{Khar}. Thus, the "spots" on the transverse plane, where the color charge is essentially nonzero have the characteristic area $s_{\perp}=\pi a^2=\frac{\pi}{Q_s^2}\approx 0.09 fm^2$. Since the transverse size of the baryonic slabs is much larger (about $\sigma_{NN}$) this means that many string-like configurations(flux tubes) are stretched between the receding slabs. Each such configuration connects the spot of opposite charges, $Q_i$, like in capacitor. In Abelian approximation the chromoelectric field strength in a flux tube is obtained from the Gauss theorem \cite{Wil},
\b
E_i=\frac{Q_i}{s_{\perp}}\equiv\rho_i
\e
Then the force acting between opposite spots is
\b
F_i=\frac12Q_iE_i=\frac12\rho_i^2s_{\perp}=s_{\perp}\epsilon_i,
\label{force}
\e 
where $\epsilon_i$ is the energy density of the chromofield. It is important to note, that all flux tubes produce attractive force between opposite spots on the projectile and target slabs. Now let us divide the slabs into $n$ small elements with equal transverse area $s_{\perp}$ so that they cover the total slab area $\sigma=ns_{\perp}$. The total force acting on each slab is then
\b
F=\frac12s_{\perp}\sum\limits_{j=1}^{n}\rho_i^2
\e
Accordingly, we can represent the integral in eq.\ref{MV} as the sum over all elements, so that 
\b
P(\rho_1,\ldots,\rho_n)\sim\exp\l(-\frac{s_{\perp}}{2}\sum\limits_{i=1}^{n}\frac{\rho_i^2}{\mu_i^2}\r)=\prod\limits_{i=1}^{n}\exp\l(-\frac{s_{\perp}\rho_i^2}{2\mu_i^2}\r)
\e  
Obviously the distribution of charge in each element follows the Gaussian distribution
\b
P(\rho_i)=\sqrt{\frac{s_{\perp}}{2\pi\mu_i^2}}\exp\l(-\frac{s_{\perp}\rho_i^2}{2\mu_i^2}\r)
\e
Therefore the mean force acting between two spots, as it follows from eqn.\ref{force}, is 
\b
<F_i>=\frac{\mu_i^2}{2}
\e

The next step is to calculate the distribution of the total force acting between the slabs for an ensemble of events. This distribution is obtained by integrating $\delta(F-\sum\limits_iF_i)$ over all charge densities $\rho_i$ with weight $P(\rho_i)$. for simplified case $\mu_1^2=\ldots=\mu_n^2=\mu^2$ we have:
$$
w(F)=\prod\limits_{i=1}^{n}\l(\int\limits_{-\infty}^{\infty}P(\rho_i)d\rho_i\r)\delta\l(F-\frac12s_{\perp}\sum\limits_{k=1}^{n}\rho_k^2\r)
$$
\b
\label{MVd}
=\l(\frac{s_{\perp}}{2\pi\mu^2}\r)^{n/2}\int\delta\l(F-\frac{s_{\perp}\rho^2}{2}\r)\exp\l(-\frac{\rho^2}{\mu^2}\r)\frac{2\pi^{n/2}}{\Gamma(n/2)}\rho^{n-1}d\rho=\frac1{\Gamma(n/2)\mu^2}\l(\frac{F}{\mu^2}\r)^{n/2-1}e^{-\frac{F}{\mu^2}}
\e
In the second expression we have used $O(4)$ symmetry of the inegrand and made transformation to spherical coordinates in n-dimensional $\rho$ space. As the result we get a gamma-distribution which has the following first moments:
$$
<F>=\frac{n}{2}\mu^2
$$
\b
\sigma_F=\sqrt{\frac{n}{2}}\mu^2
\e
We see that the parameter $\mu^2$ introduced in refs. \cite{McL,Kov}in fact determines the mean force between the slabs and its dispersion. It is more convinient to express the mean energy density of the chromofield between the slabs:
\b
\epsilon_f=\frac{<F>}{\sigma}=\frac{\mu^2}{2s_{\perp}}
\e 

Taking into account that hard gluons and valence quarks act as source of chromofield we can calculate mean squared deviation of color charge per unit area of nucleus $a$ as 
\b
\mu^2=\mu_q^2+\mu_g^2.
\e   
Quark color charge squared in a tube of transverse area $d^2{\bf s}$ is the color charge squared per unit quark $g^2C_F$ times  the number of quarks in the tube $dn_q=N_cN_a({\bf s},{\bf b})d^2{\bf s}$   
\b
\mu_q^2=g^2C_F\frac{dn_q}{d^2{\bf s}},
\e 
where $N_a({\bf s},{\bf b})$ is the number of participants from nucleus $a$ at the given radius vector ${\bf s}$ and impact parameter ${\bf b}$.
Gluon color charge squared in a tube of transverse area $d^2{\bf s}$ is the color charge squared per gluon $g^2N_c$ times  the number of gluons in the tube:   
\b
dn_g=\l\{N_a({\bf s},{\bf b})\int\limits_{x_0}^1G(x,Q_s^2)dx\r\}d^2{\bf s},
\e  
where lower integration limit is defined from relation between Bjorken variable $x$ and c.m. beam energy $\sqrt{s}$:
\b
x_0\approx\frac{2Q_0}{\sqrt{s}},
\e  
where $Q_0$ is the minimal transfered four momentum. 

Low $x$ gluon distribution function $G(x,Q^2)$ is governed by the DGLAP equation \cite{lenz}. In so-called double leading log approximation, where only terms proportional to $ln\frac1x lnQ^2$ are taken, it has the form 
\b
G(x,Q^2)=\frac1{x}exp\l(\sqrt{\frac{48}{11-\frac2{N_c}N_f}ln\l(\frac{ln\frac{Q^2}{\Lambda^2}}{ln\frac{Q_0^2}{\Lambda^2}}\r)ln\frac1{x}}\r)+\ldots
\e 
where the value $Q_0^2=1GeV^2$ was taken as a starting scale for $Q^2$ evolution. 

Running coupling constant is defined as $g_s(Q^2)=\sqrt{4\pi\alpha_s(Q^2)}$, where 
\b
\alpha_s(Q^2)=\frac{4\pi}{(11-\frac2{N_c} N_f)ln\frac{Q^2}{\Lambda^2}}
\e 
is the fine structure constant.

Saturation scale inside nucleus $a$ is obtained by the iterative solution of the following equation 
\b
Q_s^2=\frac{8\pi^2N_c}{N_c^2-1}\alpha_s(Q_s^2)N_a(b,s)\int\limits_{x_0}^1G(x,Q_s^2)dx
\e 
Therefore in dense regime number of gluons per unit area is
\b
\frac{dn_g}{d^2{\bf s}}=\frac{C_FQ_s^2}{4\pi^2\alpha_s}
\e  

Finally we have the following expression for mean color charge squared:
\b
\mu^2=C_FN_c\l(4\pi\alpha_sN_a+\frac{Q_s^2}{\pi}\r)
\e 

\section{Probability of string decay}

In this section we calculate total probability of string decay. Complicated further evolution of cascade related with numerous branchings of secondaries is out of scope of our consideration. Due to the three possible interactions which follow from ${\cal L}_{QCD}$ string can decay into quark-antiquark, gluon pair or tree gluons. Then the total density of probability is 
\b
W_{tot}(p,z,L)=W_{q\bar q}(p,z,L)+W_{gg}(p,z,L)+W_{ggg}(p,z,L)
\e  
Last term will be disregarded due to lack of analytic solution for 3-body problem. 

The density of probability of gluon pair production is calculated by the formula  
\b W_{gg}(p,z,L)=\nu_g\check{W}_g(p,z,L), \e where $\nu_{g}=2(N_c^2-1)$; $\check{W}_g=\Psi^*\Psi$, and $\Psi$ is solution of Klein-Gordon equation for massless particles (\ref{solf}). 

The density of probability of quark-antiquark pair production is defined as
\b W_{q\bar q}(p,z,L)=\nu_q\check{W}_{q\bar q}(p,z,L), \e where $\nu_{q}=N_cN_f$; $\check{W}_{q\bar q}=\Psi^{\dagger}\Psi$, and $\Psi$ is solution of Dirac equation (\ref{7}). 


Solution of the Dirac equation can be represented by the following series:
\b\Psi(x)=\sum\limits_{p}\tilde\Psi\exp[i(p_xx+p_yy-Et)],\e
where
\b\label{wave}\tilde\Psi=\sum\limits_{r=1,2}\Psi_r.\e
Eigenvectors $\Psi_r$ are Dirac spinors corresponding to the different spins \cite{ternov}: 
\b \label{spinor}
\Psi_r=a_r\l(\begin{array}{c}\psi_1+A_r\psi_2\\B_r(\psi_1+A_r\psi_2)\\A_r\psi_2-\psi_1\\-B_r(A_r\psi_2-\psi_1)\end{array}\r),
\e 
making of the eigenfunctions of the squared Dirac equation: 
\b
\label{7} \l(\l(i\frac{\partial}{\partial
t}+A_0\r)^2-\l(i\frac{\partial}{\partial r}\r)^2-m^2\pm i\sigma\r)\psi_r=0.
\e 
Cylindrical boundary conditions applied to the bag surface discretize transverse momentum inside the bag. The ground state of transverse momentum $p_{\perp}^0=c_1/r_0$ is included to effective parton mass $m_t$ \cite{wilets}; $m_t=\sqrt{m^2+(p_{\perp}^0)^2}$, where $m$ is the current parton mass; $c_1=1.4347$. 
The spin coefficients were calculated in \cite{ternov}: 
\b A_{1,2}=\frac{m_\perp\pm
ip_{\perp}}{m_{\perp}},\qquad{B_{1,2}=\pm
i\frac{p_x+ip_y}{p_{\perp}}}, \e where
$m_{\perp}=\sqrt{m_t^2+p_{\perp}^2}$;
$p_{\perp}=\sqrt{p_x^2+p_y^2}$. Therefore the decomposition
(\ref{wave}) in accordance with (\ref{spinor}) is transformed to:
\b\label{n1}\tilde\Psi=\mu_1\psi_1+\mu_2\psi_2,
\e where $\mu_1=2a_1\l(\ba{c}1\\0\\1\\0\ea\r)$;
$\mu_2=2a_2\l(\ba{c}\frac{m_t}{m_{\perp}}\\\frac{p_x+ip_y}{m_{\perp}}\\-\frac{m_t}{m_{\perp}}\\\frac{p_x+ip_y}{m_{\perp}}\ea\r)$

Spinor normalization constants are defined from condition
$\mu_{r}^{\dagger}\mu_{r}=1$: $a_{r}=\frac1{2\sqrt{2}}$, $r=1,2$.

After separation of variables (\ref{7}) takes Schrodinger-type
form: \b \label{sys}\l(\frac{\partial^2}{\partial
z^2}+p_{r}^2\r)\psi_r=0,\e where \b
p_{r}^2=\l\{\ba [c]{l}p_L^2\quad\mbox{for $z\le0$ (region I)}\\
(\omega(p)+\sigma z)^2-m_{\perp}^2\pm i\sigma
\quad\mbox{for $0\le z\le L$ (region II)} \\
\mbox{$p_R^2\quad$for$\quad z\ge L\quad$(region III)},\ea \r.\e where
$E_{L}=\omega(p)$; $E_R=\omega(p)+\sigma L$; $p_{R,L}=\sqrt{E_{R,L}^2-m_{\perp}^2}$ and
$\omega(p)\le-m_{\perp}$. Solutions in regions I, III are the following: \b\label{other}\psi_r=\l\{\ba
[c]{l}\mbox{$Ie^{ik_Ir}+R(L)e^{ik_Rr}\quad$for$\quad
z\le 0\quad$ (region I)}\\
\mbox{$T(L)e^{ik_Tr}\quad$for$\quad z\ge L\quad$ (region
III)},\\\ea\r.\e where $R,T$ are amplitudes of the reflected, and transmitted waves, respectively and four-momenta in this regions are
\b
k_I=(E_L,p_x,p_y,-p_L)
\e
\b
k_R=(E_L,p_x,p_y,p_L)
\e
\b
k_T=(E_R,p_x,p_y,p_R)
\e
\b
kr=E-p_xx-p_yy-p_zz
\e
Solution in the region II is represented as a sum of the linear independent parabolic cylinder function of order
$\nu_r=i\rho_r-1/2$; $\rho_r=-\frac{m_{\perp}^2}{2\sigma}\pm s$;
$s=\frac{i}{2}$ \cite{bateman}:
\b\label{solf}\psi_r=A(L)D_{\nu_r}^0(e^{i\frac{\pi}{4}}\zeta(p,z))+B(L)D_{\nu_r}^1(e^{i\frac{\pi}{4}}\zeta(p,z))\e where
$\zeta(p,z)=\sqrt{\frac2\sigma}\l(\omega(p)+\sigma z\r)$.
Coefficients $A(L),B(L)$ are found by matching the wave function at the boundaries.
The boundary conditions implement the wave function continuity at $z=0$ and $z=L$:
\b\l\{\ba[c]{l}(I+R(L))\mu_{r}=(A(L)D_L^0+B(L)D_L^1)\mu_{r}\\
ip_L(-I+R(L))\mu_{r}=(A(L)D_L^{0'}+B(L)D_L^{1'})\mu_{r}\\
T(L)e^{ip_RL}\mu_{r}=(A(L)D_R^0+B(L)D_R^1)\mu_{r}\\
ip_RT(L)e^{ip_RL}\mu_{r}=(A(L)D_R^{0'}+B(L)D_R^{1'})\mu_{r},\\\ea\r.\e
where $D_{L,R}^{0,1}=D^{0,1}_{\nu_r}(E_{L,R}e^{i\frac{\pi}{4}})$,
and amplitude of incident wave is $I=1$. Therefore \b
B(L)=\frac2{q_1}\frac{D_R^{0'}-ip_LD_R^0}{ip_R(D_R^1-qD_R^0)+qD_R^{0'}-D_R^{1'}};\e
\b A(L)=\frac2{q_1}-B(L)q;\e\b
R(L)=B(L)(D_L^1-qD_L^0)+\frac2{q_1}D_L^0-1;\e\b \label{other1}
T(L)=e^{-ip_RL}\l[B(L)(D_R^1-qD_R^0)+\frac2{q_1}D_R^0\r],\e
where \b q_1=D_L^0-\frac{D_L^{0'}}{ip_L};
q=\frac{D_L^1-\frac{D_L^{1'}}{ip_L}}{q_1}\e

Gluon production is described by Klein-Gordon equation: \b \label{77}
\l(\l(i\frac{\partial}{\partial t}+A_0\r)^2-
\l(i\frac{\partial}{\partial r}\r)^2\r)\Psi=0 \e 
Solution of this equation is the following: \b \label{wave1}
\Psi=\sum\limits_{p}\sum\limits_{\lambda=\pm1}\tilde\Psi\exp[i(p_xx+p_yy-Et)] e_{\lambda},
\e where $e_\lambda$ polarizations of vector particles. 

After separating of variables (\ref{77}) takes the Schrodinger-type form: \b
\l(\frac{\partial^2}{\partial z^2}+p^2\r)\tilde\Psi=0,\e
where \b
p^2=\l\{\ba [c]{l}p_L^2\quad\mbox{for $z\le0$ (region I)}\\
(\omega(p)+\sigma z)^2-m_{\perp}^2
\quad\mbox{for $0\le z\le L$ (region II)} \\
\mbox{$p_R^2\quad$for$\quad z\ge L\quad$(region III)},\ea \r.\e 
and $m_\perp=\sqrt{(\frac{c_1}{r_0})^2+p_\perp^2}$ is the gluon transverse mass.
In the region II solution of this equation is
\b\label{solf}\tilde\Psi=A(L)D_{\nu}^0(e^{i\frac{\pi}{4}}\zeta(p,z))+B(L)D_{\nu}^1(e^{i\frac{\pi}{4}}\zeta(p,z))\e where $\rho=-\frac{m_{\perp}^2}{2\sigma}$.

\section{Transmission amplitude of pair production in infinite volume}

In this section we will show that in infinite volume occupied by field the density of probability or transmission amplitude has the proper Schwinger asymptotic. Transmission amplitude is expressed by the formula (\ref{other1}) which is transformed to the following 
\b\label{trans}
T(L)\!\!=\!\!\frac{2p_L(D_R^0D_R^{1'}\!\!-\!\!D_R^{0'}D_R^1)e^{-ip_RL}}{p_L(D_R^{1'}D_L^0\!\!-\!\!D_R^{0'}D_L^1)\!\!+\!\!i(D_R^{1'}D_L^{0'}\!\!-\!\!D_R^{0'}D_L^{1'})\!\!+\!\!p_RD_R^1(D_L^{0'}\!\!-\!\!ip_LD_L^0)\!\!+\!\!p_RD_R^0(ip_LD_L^1\!\!-\!\!D_L^{1'})}
\e 
At large arguments parabolic cylinder function \cite{bateman} takes the following form $D_{R}^{0}\to e^{-\frac{\pi m_{\perp}^2}{2\sigma}}$, $D_{R,L}^{0,1'}\to\pm ip_{R,L}D_{R,L}^{0,1}$. Therefore asymptotic of transmission amplitude reads:\b
T(L)e^{ip_RL}\sim\frac{-4ip_Lp_RD_R^0D_R^1}{-2ip_Lp_RD_R^0D_L^1}=2\frac{D_R^1}{D_L^1}\sim e^{-\frac{\pi m_{\perp}^2}{2\sigma}}.\e

From latter it follows that amplitude of pair production in infinite field volume approaches the following limit: 
\b \label{4}
\varpi=\lim\limits_{L\to\infty}|T(L)|^2\sim\exp\l(-\frac{\pi m_{\perp}^2}{\sigma}\r).
\e 

\section{Partons rapidity distribution and expanding chromofields}

In the initial stage of reaction all the kinetic energy lost by baryons is transformed to chromofield:
\b
E_f(b)=\int d^2s(\epsilon_f(b,s)+B)\bar z(b,s),
\e 
where average string length is \b\label{mean} \bar z(b,s)=\sum\limits_{L=L_{min}}^{\infty}\int\limits_{-\infty}^{\infty} d^3p\int\limits_{-\infty}^{\infty}dzzW_{tot}(p,z,L),\e where $L_{min}=\frac{2m_{\perp}}{\sigma}$.  
On the other hand finally all chromofield energy is transformed to partonic plasma
\b
E_f(b)=\int d^3p\Gamma(p)=\int d^3p\omega(p)\frac{dN}{d^3p}
\e 
Therefore rapidity distribution of partons generated in the slab-slab collision is calculated by the formula: 
\b
\frac{dN}{dy}(b)=\int\limits_{-\infty}^{\infty}d^2p_{\perp}\Gamma(p)
\e

\section{Conclusions}

\begin{figure}
\includegraphics[width=8.6cm]{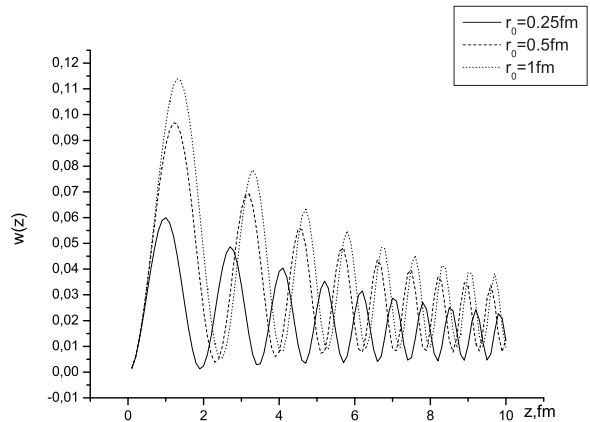}
\caption{Density of probability of quark-antiquark pair production
$W_{q\bar q}^f(z,p_{\perp},y,L)$ as a function of coordinate inside the color flux tube of length $L=10 fm$ 
at the fixed rapidity $y=0.21$ and average transverse momentum $<p_{\perp}>=0.5 GeV$.} \label{fig2}
\end{figure}

The results of this article are following: 1)the total density of probability of partons production is obtained by the
exact solutions of squared Dirac and Klein-Gordon equations in the linear vector potential; 2)Rapidity spectra of the field energy and produced particles are calculated from the energy conservation law; 3)It was shown that the density of probability of particles production in the infinite field volume approaches the classical Schwinger's result. 

Numerical simulations were performed for most central $Au+Au$ collisions with c.m.energy
$\sqrt{s}=200 AGeV$. The coordinate dependencies of a $q\bar{q}$--pair production
density of probability at the different color flux tubes radii ($r_0=0.25fm,0.5fm,1fm$) are shown on Fig.\ref{fig2}. Quarks have current mass $m_q\sim 8 MeV$ and gluons are assumed massless.

\acknowledgments{ The author wishes to express sincere appreciation to Prof. I.N. Mishustin for many insightful discussions.}

\end{document}